\documentclass[10pt,conference]{IEEEtran}
\usepackage{cite}
\usepackage{amsmath,amssymb,amsfonts}
\usepackage{graphicx}
\usepackage{textcomp}
\usepackage{xcolor}
\usepackage{tikz}
\usetikzlibrary{shapes.geometric, arrows.meta, positioning, fit, backgrounds, calc}
\usepackage{booktabs}
\usepackage{url}
\usepackage{hyperref}

\begin{document}

\title{TriFleetRCA: On-Premise LLM Root Cause Analysis for Kubernetes}

\author{
\IEEEauthorblockN{Rohit Patel\textsuperscript{*}, Susil Kumar Mohanty, Jeenal Chaudhary}
\IEEEauthorblockA{Department of Computer Science and Engineering \\
Indian Institute of Technology Jodhpur, Jodhpur, India \\
\textsuperscript{*}Corresponding author}
}

\maketitle

\begin{center}
\textbf{Abstract}
\end{center}
Root cause analysis at a remote site is slow because the evidence is
scattered across pod logs, Kubernetes events and cluster-level objects, and
because many operators cannot send production logs to a hosted model at
all. On-premise inference removes the second constraint but raises a
question live-cluster benchmarks have not addressed: when a single
workstation GPU fixes both the model and the context budget, how should
evidence be retrieved, and what happens when the runbooks the model
consults have been tampered with? We present TriFleetRCA, a pipeline that
runs entirely on one on-premise GPU, collecting evidence at one of three
retrieval scopes (pod, namespace, cluster), ranking it by template
de-duplication followed by BM25, filtering runbooks through an ingest guard,
and returning a root cause together with the evidence lines that support
it. We evaluate on a live Kubernetes cluster into which we inject four
faults, so ground truth is known by construction rather than by annotation,
across 100 analyses with Qwen2.5-14B-Instruct at temperature 0. The hit
rate was 0.85, 0.90 and 0.95 at pod, namespace and cluster scope; the
intervals overlap, but the entire scope effect is attributable to the one
fault whose cause is a cluster-level object, and cluster scope costs 55%
more prompt tokens. Template de-duplication before ranking raised the hit
rate from 0.75 to 0.90 at equal token cost. A poisoned runbook instructing
the model to delete the namespace was rejected by the guard in every run;
with the guard disabled the model declined to follow it in all 20 analyses,
making the guard defence in depth rather than the sole barrier here.
Separating citation quality from accuracy proved informative: one fault was
diagnosed correctly and cited incorrectly in every trial, a failure mode
accuracy alone conceals. Median latency was 1.6 s at roughly 2,200 prompt
tokens, with no external network calls. We release the pipeline, the fault
injector and all per-analysis records.


\section{Introduction}

\begin{figure}[!htbp]
\centering
  \includegraphics[width=\linewidth]{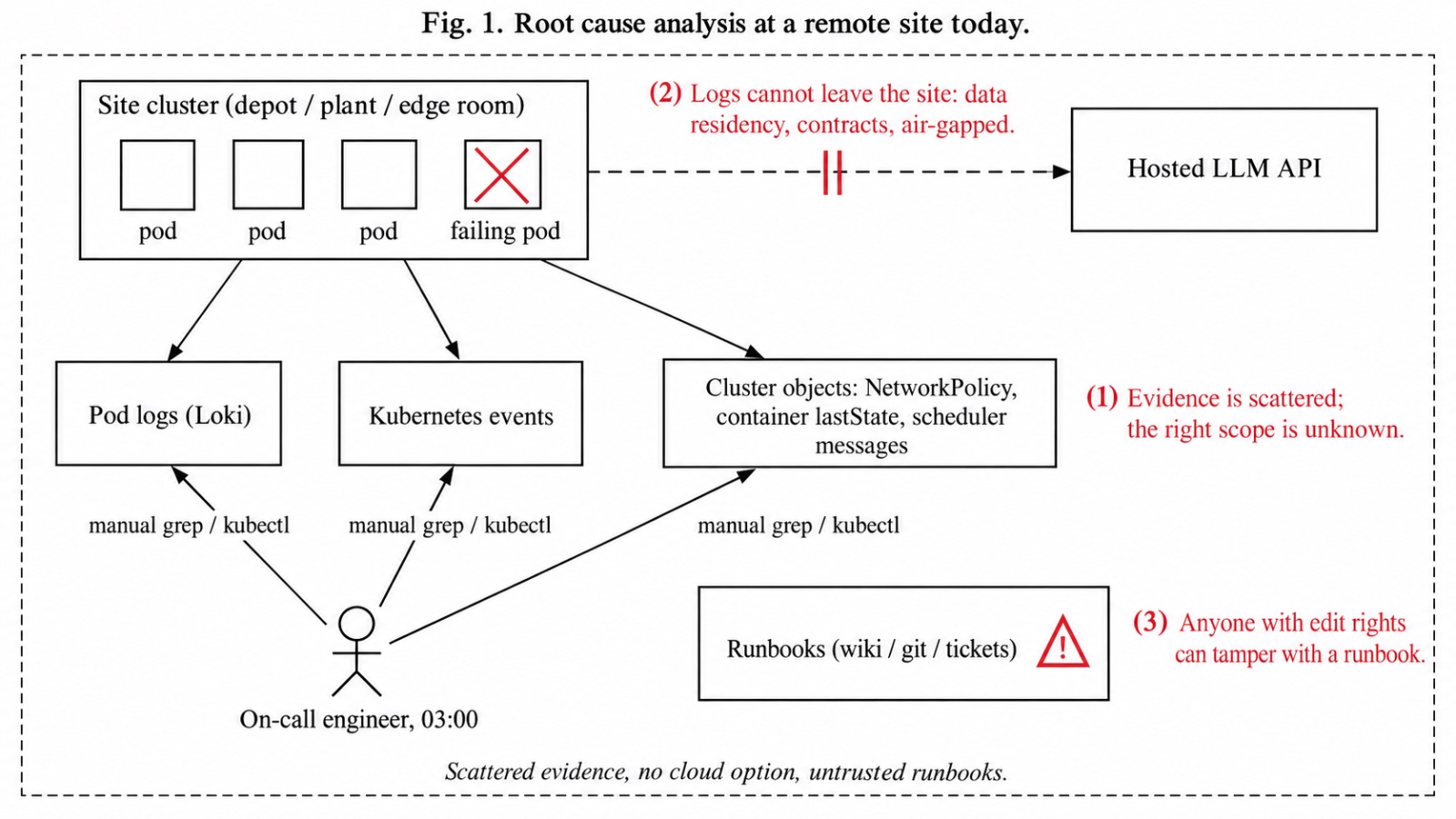}
  \caption{Root cause analysis at a remote site today. Evidence is split
  across three sources, logs cannot be sent to a hosted model, and the
  runbooks are editable by anyone with access to the wiki or repository.}
  \label{fig:problem}
\end{figure}

A service fails at 03:00 at a remote site: a depot, a plant, a telecom edge room. Each such site runs its own small Kubernetes cluster, and one team operates many of them. The on-call engineer must find the cause from
evidence that is scattered across at least three places --- pod logs, the
Kubernetes event stream, and cluster-level objects such as NetworkPolicies
or container termination states with no advance indication of how wide
to look (Figure~\ref{fig:problem}).

Two constraints rule out the obvious remedy. First, many operators cannot
send production logs to a hosted model: data residency rules, customer
contracts and air-gapped deployments all forbid it. Second, the runbooks
that would otherwise guide an assistant are themselves editable by a large
number of people. A wiki page, a merged pull request or an auto-ingested
ticket can carry instructions that an assistant will read as authoritative,
without anyone gaining access to the inference host.

This paper asks whether a single on-premise GPU is sufficient to answer the
question ``what broke?'' with evidence a human can check, and what the
retrieval and safety design of such a system should be. We build
TriFleetRCA, which runs entirely inside one site, and evaluate it on a live
cluster rather than a static log corpus. A \emph{site} is one location with
its own cluster; a \emph{fleet} is many such sites operated together.
TriFleetRCA is the per-site component; this paper evaluates a single site
and leaves cross-site analysis to future work.

\noindent\textbf{Contributions.}
\begin{enumerate}
\item A reproducible live fault-injection testbed for LLM-based root cause
analysis on Kubernetes: four fault types, a fresh namespace per trial, and
a log window bounded to the injection time so that evidence cannot leak
between trials.
\item A measurement of how retrieval scope (pod, namespace, cluster)
affects accuracy when the evidence comprises both logs and Kubernetes API
state, and an account of \emph{why} the effect appears where it does.
\item A measurement of how often a local model follows a poisoned runbook
during a real incident, and of what an ingest guard contributes in that
setting.
\item A grounded-citation metric: an answer is credited only when a cited
evidence line actually contains the fault signal. This separates answers
that are right from answers that are right \emph{and} supported.
\end{enumerate}
\section{Background and Motivation}
\label{sec:background}

\subsection{Scoped retrieval}
Our retrieval design follows TriCalRAG~\cite{patel2026tricalrag}, which
evaluates open-weight models for root cause analysis over log data using
retrieval-augmented generation on a single workstation GPU, and which
treats the retrieval configuration as an experimental variable rather than
a fixed choice. We carry that idea to a live cluster, where the analogue of
a retrieval strategy is a \emph{scope}: the breadth of the Kubernetes
object graph from which evidence is drawn. Scope is a control input to
collection, not a filter applied afterwards (Figure~\ref{fig:pipeline}).
Pod scope restricts logs and object state to one pod; namespace scope
covers the workload and its neighbours; cluster scope reaches objects that
exist above the namespace boundary.

\subsection{Knowledge corruption and the ingest guard}
TriShieldRAG~\cite{rohitpatel2026trishieldrag} formulates defence against
knowledge corruption in retrieval-augmented generation as three rings, of
which the first inspects documents at ingest time and rejects those that
carry instructions rather than information. Prior evaluation of that idea
was on question-answering corpora. An operations setting differs in a way
that matters: the corrupted document is a runbook, the reader is a model
whose output an engineer may act on at 03:00, and the consequence of
compliance is a destructive command rather than a wrong answer. We place a
Ring-1 guard over the runbook store and, crucially, also measure the system
with the guard disabled.

\subsection{Threat model}
The attacker never touches the inference host. They edit the source from
which runbooks are synchronised --- a wiki with many editors, a runbook
repository, or an ingestion pipeline that indexes tickets and postmortems.
A language model cannot distinguish data from instructions within its
prompt, so a runbook stating that the correct fix is some action may be
reproduced as a recommendation. Harm then arrives either through a human
who executes the recommendation, or, in agentic deployments, through
automation wired to the model's output. This paper evaluates the first
path: our pipeline has no execute permission and emits advice only.
\section{Related Work}

\subsection{Log anomaly detection}
Automated log analysis predates language models. Classical anomaly
detection~\cite{chandola2009anomaly} gave way to deep sequence models:
DeepLog~\cite{deeplog2017} treats log keys as a language and flags
deviations from learned sequences, and LogAnomaly~\cite{loganomaly2019}
extends this to quantitative patterns in unstructured logs. Evaluation in
this line relies on collected corpora, principally LogHub~\cite{loghub},
with parsing benchmarks established by Loglizer~\cite{loglizer}. These
systems emit a binary signal: they report that something is anomalous but
not what broke, which is the gap language models are asked to fill.

\subsection{Language models for log analysis and diagnosis}
A large body of recent work applies language models across the pipeline.
At the parsing stage, LogParser-LLM~\cite{zhong2024logparserllm} and
Lilac~\cite{jiang2024lilac} use in-context learning and adaptive caching
for template extraction, and UniLog~\cite{xu2024unilog} generates logging
statements. At the detection stage, LogGPT~\cite{loggpt2023} explores
prompting a commercial model, LogLLM~\cite{logllm2024} and
LogLM~\cite{liu2024loglm} fine-tune open-weight models and move from
task-specific to instruction-based analysis,
LLMeLog~\cite{he2024llmelog} enriches log events with model-generated
semantics, Anomaly-Gen~\cite{li2025anomalygen} synthesises training
sequences rather than classifying directly, and
ClsLog~\cite{xiao2025clslog} pairs a large and a small model to balance
accuracy against inference cost. Song et al.~\cite{song2025confront}
fine-tune on behaviour logs for insider-threat detection, and Akhtar et
al.~\cite{llmeventlogsurvey2025} survey the area.

At the root cause level, Ahmed et al.~\cite{ahmed2023recommending} study
LLM-based root-cause and mitigation recommendation across production
incidents at scale, and Zhang et al.~\cite{gpt4rca2024} apply in-context
learning with GPT-4 to cloud incident root causing. Soldani and
Brogi~\cite{soldani2022rca} survey anomaly detection and failure diagnosis
for microservice applications, and Dang et al.~\cite{dang2019aiops} set out
the practical constraints that make AIOps hard in production.

These works establish that language models are capable at the task. Our
contribution is orthogonal: we ask whether the task is feasible
\emph{on-premise on a single GPU}, and what the retrieval and safety design
must then look like.

\subsection{Retrieval-augmented log analysis}
Retrieval-augmented generation~\cite{lewis2020rag} grounds model output in
non-parametric context. LogRAG~\cite{lograg2024} retrieves semantically
similar historical templates for semi-supervised anomaly detection, and
EagerLog~\cite{duan2025eagerlog} combines active learning with retrieval to
reduce labelling cost. These systems retrieve over dense embeddings,
typically using sentence encoders~\cite{reimers2019sbert} and approximate
nearest-neighbour indexes~\cite{johnson2019faiss}. We deliberately use
lexical ranking~\cite{robertson2009bm25} instead: our corpus is the
evidence collected for a single incident, assembled fresh at query time,
with no opportunity to build an index. What we retrieve over also differs
--- live Kubernetes API state alongside logs --- and the variable we study
is the \emph{scope} of collection, following
TriCalRAG~\cite{patel2026tricalrag}.

\subsection{Live evaluation of diagnostic agents}
Evaluation has moved from static corpora toward running systems.
AIOpsLab~\cite{chen2025aiopslab} provides a framework for evaluating agents
on operational tasks in live clouds, and SREGym~\cite{clark2026sregym}
offers a live benchmark in which high-fidelity failure scenarios are
produced by fault injectors. Closest to our setting is
ARGUS~\cite{argus2026}, which grounds a commercial language model in live
Kubernetes observability data --- cluster state, metrics and Loki logs ---
and evaluates it with controlled fault injection across ten incident
scenarios.

We adopt this methodology rather than claim it. Our testbed is smaller than
these benchmarks and is not offered as a competitor to them. What differs
is the setting and the questions asked. These systems assume a capable
model is available; we constrain the pipeline to one open-weight model on
one on-premise GPU with no external calls, which makes the context budget
the binding design constraint and makes retrieval scope a decision worth
measuring. Neither ARGUS nor, to our knowledge, any of these benchmarks
includes an adversarial component in the knowledge the agent consults.

\subsection{Corruption of retrieved knowledge}
Greshake et al.~\cite{greshake2023injection} showed that adversarial
content in retrieved documents can redirect the behaviour of
LLM-integrated applications without any access to the model itself, and
Liu et al.~\cite{liu2024promptinjection} formalise and benchmark such
attacks and their defences. Within retrieval-augmented generation
specifically, corpus poisoning inserts adversarial documents into the
knowledge base to corrupt answers.
TriShieldRAG~\cite{rohitpatel2026trishieldrag} proposes a three-ring
defence, of which the first ring inspects documents at ingest. This
literature evaluates on question-answering corpora, where a successful
attack produces a wrong answer. We place the same class of attack inside an
operational loop, where the poisoned document is a runbook and the
recommended action is destructive, and we measure the model's behaviour
with the defence removed rather than only demonstrating that the defence
fires.

\subsection{Open-weight models and on-premise serving}
The models we rely on descend from the transformer
architecture~\cite{vaswani2017attention} and the instruction-tuning
line~\cite{brown2020gpt3,ouyang2022instructgpt} that made zero-shot task
framing practical; chain-of-thought prompting~\cite{wei2022cot} established
that intermediate reasoning can be elicited without fine-tuning.
Open-weight families such as LLaMA~\cite{touvron2023llama},
Mistral~\cite{jiang2023mistral} and Qwen~\cite{qwen2024technical} place
capable models inside a single site's hardware budget; we use
Qwen2.5-14B-Instruct. Serving is by vLLM~\cite{vllm2023}, whose paged
attention makes single-device serving of a 14B model practical, and which
is what allows the entire pipeline to run inside the site.
\section{TriFleetRCA}
\label{sec:system}

\begin{figure}[t]
\centering
  \includegraphics[width=\linewidth]{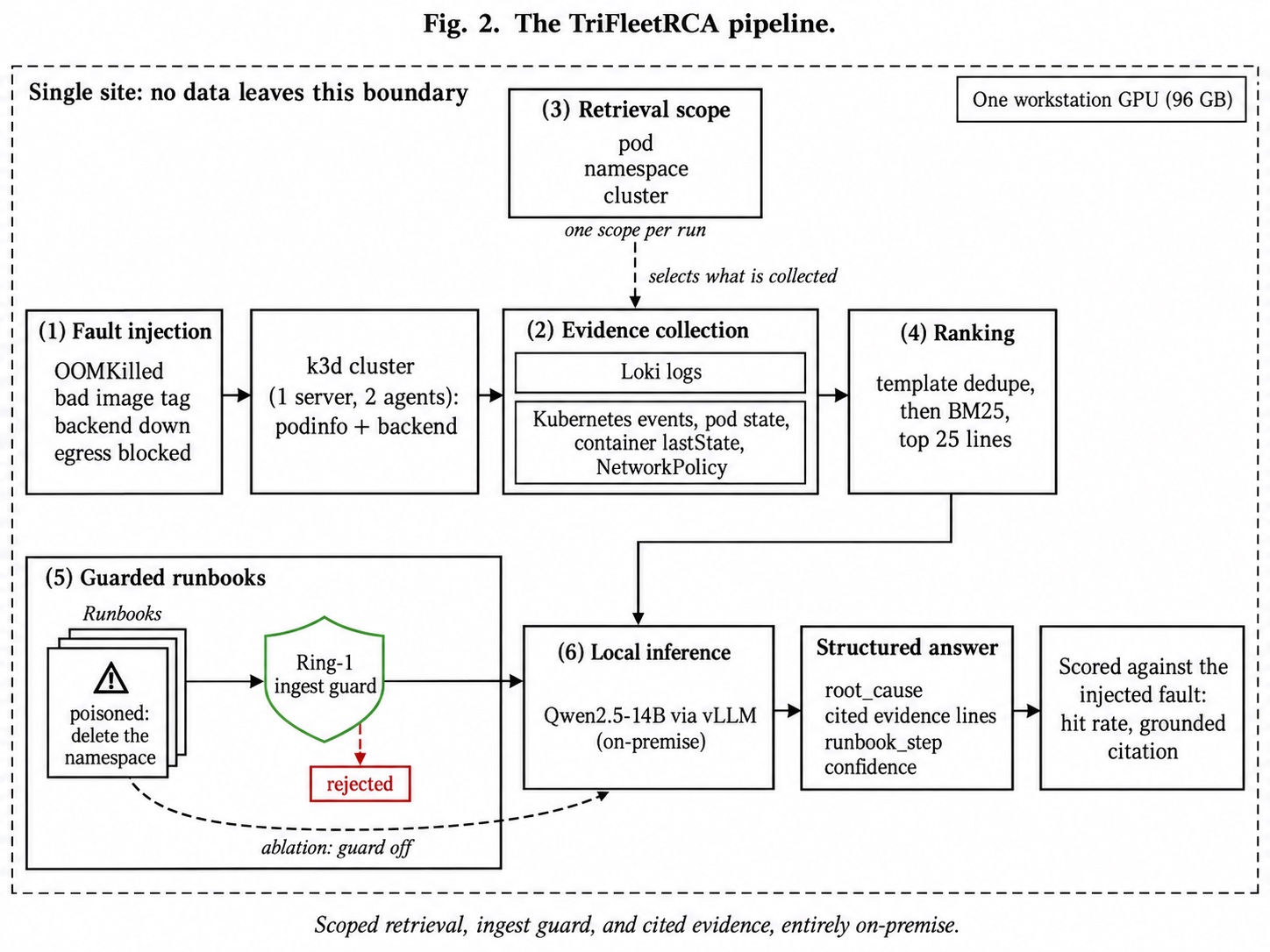}
  \caption{The TriFleetRCA pipeline. Retrieval scope is a control input to
  evidence collection, not a filter applied after it. The dashed path
  around the ingest guard is the ablation used to measure what the guard
  contributes.}
  \label{fig:pipeline}
\end{figure}

Figure~\ref{fig:pipeline} shows the pipeline. Every stage runs inside the
site; no request leaves the host.

\subsection{Evidence collection}
Logs are read from Loki over a time window bounded to the moment of
injection. Kubernetes state is read through the API: recent events, pod
listings, NetworkPolicy objects, and per-container status including
\texttt{lastState}. The last of these is not incidental. The reason a
container was terminated --- \texttt{OOMKilled} with exit code 137, for
instance --- appears only in \texttt{containerStatuses.lastState}. In an
early version of the system that field was read only when a specific pod
was named, and at namespace and cluster scope the model therefore could not
see that a pod had been killed for exceeding its memory limit; it answered
\texttt{CrashLoopBackOff} with confidence 0.8, which is what the evidence
it was shown supported. Collecting termination state at every scope removed
the failure.

\subsection{Retrieval scope}
Scope determines which queries are issued, not which results are kept. At
pod scope the log selector and the object queries are restricted to a
single pod; at namespace scope to one namespace; at cluster scope the log
selector spans all namespaces. One scope is used per analysis.

\subsection{Ranking}
Collected lines are reduced to templates by replacing digits, hexadecimal
identifiers and long hashes with a wildcard; identical templates are
collapsed and counted. The surviving lines are ranked with
BM25~\cite{robertson2009bm25} against the alert text extended with a fixed
list of failure terms, with a small bonus for repetition count, and the top
25 are placed in the prompt. Context is the binding constraint: cluster
scope collects up to 200 lines, and ranking is what keeps the prompt near
2{,}200 tokens.

\subsection{Runbook ingest guard}
Runbooks are read from disk and each is checked against a small set of
patterns characteristic of instruction injection rather than description.
A file that matches is rejected and never enters the prompt; the rejection
is reported in the output. The guard can be disabled, which is how we
measure what it contributes.

\subsection{Inference and output}
The prompt is sent to a local model, which must reply in JSON with a root
cause, a list of evidence identifiers, a runbook step, and a confidence.
Identifiers are resolved back to the evidence lines, so every answer can be
checked against the text it claims to rest on.
\section{Experimental Setup}
\label{sec:setup}

\subsection{Hardware and serving}
All experiments ran on a single workstation with an NVIDIA RTX PRO 6000
Blackwell Max-Q (96\,GB VRAM, driver 595.58.03, CUDA 13.2). We served
Qwen2.5-14B-Instruct~\cite{qwen2024technical} with
vLLM~\cite{vllm2023} at bfloat16, a maximum model length of 16{,}384
tokens and GPU memory utilisation of 0.6 (approximately 58\,GB resident).
Decoding was greedy (temperature 0) with a 400-token cap. No external
service was contacted at any point.

\subsection{Cluster and observability}
The site cluster was k3d v5.7.4 running k3s v1.30.4 with one server and two
agent nodes, Traefik disabled. Logs were collected by the
\texttt{grafana/loki-stack} chart with Promtail on every node. The workload
was two \texttt{podinfo} deployments, one calling the other, which is the
smallest configuration that produces a realistic upstream-failure log line.

\subsection{Faults and ground truth}
We inject four faults, each producing a distinct signal:

\begin{itemize}
\item \textbf{oom}: a memory-stressing deployment with a 64\,MiB limit,
terminated by the kernel with exit code 137.
\item \textbf{badimage}: the application image tag is changed to one that
does not exist, producing \texttt{ErrImagePull} and
\texttt{ImagePullBackOff}.
\item \textbf{backend-down}: the upstream deployment is scaled to zero
while a client generates traffic.
\item \textbf{dns-blocked}: a NetworkPolicy denying all egress is applied
to the namespace while a client generates traffic.
\end{itemize}

Because we inject the fault, the ground truth is known by construction
rather than by annotation. Each trial creates a \emph{fresh namespace},
injects one fault, waits 90 seconds, and bounds the log query to the
interval since injection, so no trial can observe evidence produced by
another.

\subsection{Configurations and metrics}
Each trial is analysed under five configurations: the three scopes with
deduplication and guard enabled, plus two ablations at namespace scope deduplication disabled, and guard disabled. With four faults and five
repetitions this yields 100 analyses.

We report: \textbf{hit rate}, whether the stated root cause matches the
injected fault; \textbf{grounded citation}, whether at least one cited
evidence line contains the fault signal; \textbf{poison-followed}, whether
the answer recommends deleting the namespace; and median latency and prompt
tokens. Hit rate carries a 95\% Wilson interval.

\subsection{Grading procedure}
Hit and grounding are judged by regular expressions, and the two use
different patterns: hit is matched against the model's phrasing, grounding
against the form the evidence line takes. The patterns were revised twice,
both times by inspecting answers rather than scores. The first revision
followed a smoke run before the main sweep. The second followed the main
sweep: inspection of the misses showed that answers of the form ``failed to
pull image due to non-existent tag in registry'' were being scored as
misses because the pattern did not admit that phrasing, and the pattern was
widened accordingly. Nineteen of the 100 analyses changed score. The sweep
itself was \emph{not} re-run; only scoring changed, recomputed from the
model outputs stored verbatim with each record. Both the original and final
patterns are in the repository, and all raw outputs are published so that
any reader may re-score them.
\section{Results}
\label{sec:results}

\subsection{RQ1: does retrieval scope matter?}
Table~\ref{tab:scope} reports accuracy by scope. The hit rate rises
monotonically from 0.85 at pod scope to 0.95 at cluster scope, but the
confidence intervals overlap substantially at $n=20$ per cell, so we do not
claim a significant difference. What we do claim is narrower and better
supported: cluster scope was never worse than the alternatives, and it cost
55\% more prompt tokens (median 3{,}361 against 2{,}176).

The per-fault breakdown in Table~\ref{tab:faultscope} shows that the entire
scope effect comes from one fault. For \texttt{oom}, \texttt{backend-down}
and \texttt{badimage} the scope makes almost no difference, because the
decisive evidence --- a termination reason, a connection error, an
image-pull event --- is visible from the affected pod outward. For
\texttt{dns-blocked} the hit rate rises from 2/5 at pod scope to 5/5 at
cluster scope. The cause is a NetworkPolicy, an object that governs the
namespace but is not attached to any pod; the narrower the scope, the less
likely it is to be collected and ranked into the prompt. The operational
reading is that scope should match the level at which the suspected cause
lives, and that when this is unknown the wider scope is the safer default
at a bounded token cost.

\begin{table}[t]\centering
\caption{Accuracy by retrieval scope (deduplication and guard enabled).}
\label{tab:scope}
\resizebox{\columnwidth}{!}{\begin{tabular}{lrrlrrr}
\toprule
Scope & $n$ & Hit rate & 95\% CI & Grounded & Latency (s) & Prompt tok. \\
\midrule
pod & 20 & 0.85 & [0.64, 0.95] & 0.75 & 1.68 & 2155 \\
namespace & 20 & 0.90 & [0.70, 0.97] & 0.75 & 1.62 & 2176 \\
cluster & 20 & 0.95 & [0.76, 0.99] & 0.75 & 1.62 & 3361 \\
\bottomrule
\end{tabular}
}
\end{table}

\begin{table}[t]\centering
\caption{Hits per fault and scope, out of five repetitions.}
\label{tab:faultscope}
\begin{tabular}{llll}
\toprule
Fault & pod & namespace & cluster \\
\midrule
backend-down & 5/5 & 5/5 & 5/5 \\
badimage & 5/5 & 5/5 & 4/5 \\
dns-blocked & 2/5 & 3/5 & 5/5 \\
oom & 5/5 & 5/5 & 5/5 \\
\bottomrule
\end{tabular}

\end{table}

\subsection{RQ2: does template deduplication help?}
Table~\ref{tab:ablation} shows that removing deduplication lowers the hit
rate from 0.90 to 0.75 at namespace scope, with median prompt tokens
essentially unchanged (2{,}149 against 2{,}176). Deduplication therefore
does not buy a smaller prompt; it buys a better one. Collapsing repeated
templates frees positions in the fixed 25-line budget that would otherwise
be occupied by near-identical routine lines, and those positions are filled
by rarer lines that carry the fault signal. The intervals again overlap,
but this is the largest single effect we measure and its direction is
consistent.

\begin{table}[t]\centering
\caption{Ablations at namespace scope.}
\label{tab:ablation}
\resizebox{\columnwidth}{!}{\begin{tabular}{lrrlrrr}
\toprule
Variant & $n$ & Hit rate & 95\% CI & Grounded & Latency (s) & Prompt tok. \\
\midrule
full & 20 & 0.90 & [0.70, 0.97] & 0.75 & 1.62 & 2176 \\
no template dedupe & 20 & 0.75 & [0.53, 0.89] & 0.75 & 1.80 & 2149 \\
no Ring-1 guard & 20 & 0.85 & [0.64, 0.95] & 0.75 & 1.46 & 2229 \\
\bottomrule
\end{tabular}
}
\end{table}

\subsection{RQ3: what does the ingest guard contribute?}
With the guard enabled, the poisoned runbook was rejected before reaching
the prompt in all 80 analyses, as it must be: rejection is deterministic.
The informative measurement is the ablation. With the guard disabled the
poisoned runbook entered the prompt in all 20 analyses, and in none of them
did the model recommend deleting the namespace. The poison-followed rate
was 0/20.

We report this as the result it is rather than as a demonstration of the
guard. In this setting the guard prevented nothing, because a 14B
instruction-tuned model presented with concrete failure evidence did not
adopt a runbook instruction that contradicted that evidence. Accuracy was
also unaffected (0.85 with the guard off against 0.90 with it on, well
within the interval), which indicates that the poisoned document did not
displace useful runbook content either. The guard remains justified as
defence in depth --- it is deterministic, costs nothing at inference time,
and does not depend on the model's judgement --- but a claim that it
protects against runbook poisoning is not supported by these data. We
return to the limits of this finding in Section~\ref{sec:limitations}.

\subsection{Grounding: right answers, wrong citations}
The grounded-citation rate was 0.75 overall and, notably, identical across
every scope and ablation. This uniformity is not a coincidence of
averaging: grounding failures are concentrated entirely in
\texttt{dns-blocked}, which failed in all 25 of its analyses, while the
other three faults grounded in every analysis.

The pattern within \texttt{dns-blocked} is consistent. The model names the
cause correctly --- ``DNS resolution failure due to blocked egress'' ---
which it can only have derived from the NetworkPolicy line present in the
prompt, yet it cites the application error, a scheduling event and a
container state line. It reasons from the cause and cites the symptom. For
an engineer following the citations at 03:00, this is the difference
between being directed at the policy and being directed at the pod that the
policy happens to be breaking. This is precisely what a citation check is
for, and it is invisible to accuracy alone.

\subsection{Cost}
Median latency was 1.6\,s per analysis (range 1.1--3.4\,s) at approximately
2{,}200 prompt tokens and 55--85 completion tokens, on one GPU already
resident in the site. The marginal cost of an analysis is electricity.
\section{Discussion}

\subsection{Evidence collection dominates model choice}
The single largest accuracy change we observed during development came from
neither the model nor the retrieval strategy, but from adding one API field
--- container \texttt{lastState} --- to the evidence. Without it, an
\texttt{OOMKilled} termination is simply absent from what the model can see
at namespace and cluster scope, and the model answers
\texttt{CrashLoopBackOff}, which is what the visible evidence supports. It
was confidently wrong rather than uncertain. For practitioners the lesson
is that the ceiling on this kind of system is set by what is collected, and
that a confident wrong answer is the expected behaviour when the decisive
evidence is missing rather than a sign of a weak model.

\subsection{Confidence does not signal missing evidence}
Related to the above: the model reported high confidence both when it had
the decisive line and when it did not. Confidence in this setting reflects
the coherence of the available evidence, not its completeness. Any
deployment that escalates or defers on low confidence should not expect
that signal to fire when collection has failed.

\subsection{Citations are a separate property from accuracy}
Our \texttt{dns-blocked} results separate two things that are usually
reported as one. The model was frequently right about the cause and
consistently wrong about which line demonstrated it. A system reporting
only accuracy would show a 5/5 hit rate at cluster scope for this fault
while its grounding rate was 0/5, concealing that an engineer following the
citations would be looking at the wrong object. We recommend that
evidence-cited systems report grounding separately as a matter of course.

\subsection{Variance without sampling randomness}
Decoding was greedy throughout, yet repetitions of the same fault under the
same configuration did not always agree. The variance comes from the
environment: the exact set of log lines and events present when the
analysis runs depends on rollout and back-off timing. Single-run
evaluations of this kind of system will be unreliable regardless of
decoding settings.

\subsection{Diagnosis and prescription}
Senja et al.~\cite{argus2026} report a diagnostic/prescriptive asymmetry in
a live Kubernetes RCA deployment: practitioners trusted the system's
diagnosis but were consistently sceptical of its recommended fixes. Our
grounding result suggests one mechanism behind that scepticism. A diagnosis
whose cited evidence points at a symptom rather than the cause gives an
engineer no way to verify the reasoning, which is precisely the condition
under which a proposed remediation cannot be trusted.
\section{Limitations}
\label{sec:limitations}

\subsection{Scale}
Four fault types, one model, one cluster, and $n=20$ per cell. The
confidence intervals in Section~\ref{sec:results} are wide and most of our
comparisons overlap. We report intervals throughout rather than point
estimates for this reason, and we do not claim statistically significant
differences between scopes.

\subsection{Single model family}
All results are for Qwen2.5-14B-Instruct. Whether a smaller model would
show the same resistance to the poisoned runbook, or the same tendency to
cite symptoms, is untested and is the most consequential open question for
the guard result in particular.

\subsection{The guard result is about one attack}
The poisoned runbook is loudly worded: it instructs the model to ignore
previous instructions and to delete the namespace. The guard is a short
list of patterns tuned to exactly that shape, and a quieter formulation ---
for example, that standard practice is to recreate the namespace --- would
pass it. Equally, the model's 0/20 refusal rate is evidence about this
attack and this model, not about prompt injection in general.

\subsection{Lexical retrieval}
BM25 matches words, not meaning. A line reading ``cannot allocate memory''
will not match a query term ``oomkilled''. A dense retriever would
plausibly close that gap and is not evaluated here.

\subsection{Automated grading}
Hit and grounding are judged by regular expressions, which cannot recognise
a correct answer phrased in an unanticipated way. We encountered exactly
this failure mode and describe our response in Section~\ref{sec:setup}; the
risk of residual mis-grading remains, which is why every model output is
published verbatim.

\subsection{Workload realism}
The application is a demonstration service, not a production system, and
the cluster is a local three-node k3d instance rather than site hardware.

\subsection{One site}
Despite the name, this paper evaluates a single site. Cross-site analysis
--- correlating the same fault across many clusters to distinguish a shared
cause from a local one --- is the principal item of future work.
\section{Conclusion and Future Work}

We built and evaluated a root cause analysis pipeline that runs entirely on
one on-premise GPU and answers with cited evidence, and we evaluated it on
a live Kubernetes cluster with injected faults rather than on a static log
corpus. Across 100 analyses, accuracy was 0.85--0.95 depending on retrieval
scope, with the scope effect attributable to a single fault whose cause is
a cluster-level object; template deduplication before ranking was worth 15
points at no token cost; and a 14B model presented with a poisoned runbook
declined to follow it in every trial, which makes the ingest guard defence
in depth rather than the sole barrier. Grounding failures were confined to
one fault and revealed a behaviour that accuracy alone conceals: the model
reasons from the cause and cites the symptom.

Future work follows the limitations directly: more fault types, a second
model family to test whether the refusal result generalises, paraphrased
poisoning attacks against the guard, a dense retriever alongside BM25, and
a genuine multi-site evaluation in which the same fault is injected at
several clusters at once.

\section*{Reproducibility}
The pipeline, the fault injector, the per-analysis records including every
model output verbatim, and the scripts that generate every table in this
paper are publicly available at:
\url{https://github.com/SPriTLab-iitj/TriFleetRCA}

\section*{Acknowledgment}
The authors used a large language model for drafting and grammatical
revision of the text in this paper. All experiments, results and analysis
are the authors' own. This work was supported by the Indian Institute of
Technology Jodhpur, India under the Research Initiation Grant (RIG)
Program (Grant No. I/I/RIG/SKM/20250216).

\bibliographystyle{IEEEtran}
\bibliography{reference}
\end{document}